\documentclass[twocolumn,showpacs,preprintnumbers,amsmath,amssymb]{revtex4}
\usepackage{amsmath}

\usepackage{float}
\usepackage{graphicx}
\usepackage{dcolumn}
\usepackage{bm}
\usepackage{color}
\definecolor{orange}{RGB}{255,127,0}
\definecolor{brown}{RGB}{160,82,45}

\begin{document}

\def\vsigma{{\hbox{\boldmath $\sigma$}}}
\def\bp{{\hbox{\boldmath $p$}}}
\def\bq{{\hbox{\boldmath $q$}}}
\def\undersim#1{\setbox9\hbox{${#1}$}{#1}\kern-\wd9\lower
    2.5pt \hbox{\lower\dp9\hbox to \wd9{\hss $_\sim$\hss}}}

\title{Pion transverse momentum spectrum, elliptic flow and interferometry in the granular source model
for RHIC and LHC heavy ion collisions}

\author{Jing Yang$^1$}
\author{Yan-Yu Ren$^2$}
\author{Wei-Ning Zhang$^{1,\,2}$\footnote{wnzhang@dlut.edu.cn}}
\affiliation{$^1$School of physics and optoelectronic technology, Dalian University of Technology,
Dalian, Liaoning 116024, China\\
$^2$Department of Physics, Harbin Institute of Technology, Harbin, Heilongjiang 150006, China}

\begin{abstract}
We systematically investigate the pion transverse momentum spectrum, elliptic flow, and Hanbury-Brown-Twiss
(HBT) interferometry in the granular source model for the heavy ion collisions of Au-Au at $\sqrt{s_{NN}}=$
200 GeV and Pb-Pb at $\sqrt{s_{NN}}=$ 2.76 TeV with different centralities.  The granular source model can
well reproduce the experimental results of the heavy ion collisions at the Relativistic Heavy Ion Collider
(RHIC) and the Large Hadron Collider (LHC).  We examine the parameters involved in the granular source model.
The experimental data of the momentum spectrum, elliptic flow, and HBT radii for the two collision energies
and different centralities impose very strict constraints on the model parameters.  They exhibit certain
regularities for collision centrality and energy.  The space-time structure and expansion velocities of the
granular sources for the heavy ion collisions at the RHIC and LHC energies with different centralities are
investigated.
\end{abstract}


\maketitle

\section{Introduction}

The main purpose of relativistic heavy ion collisions is to probe the new matter, quark-gluon plasma (QGP),
and study its properties.  Because of the complexity of the nucleus-nucleus collisions, model investigation
plays important roles in determining and characterizing the QGP.  Single particle spectrum, elliptic flow,
and two-particle Hanbury-Brown-Twiss (HBT) correlations are crucial final-state observables in relativistic
heavy ion collisions \cite{{STA-spe04z,PHE-spe04z,PHO-spe07,STA-v2-02,PHE-v2-02,PHE-v2-03,STA-v2-05z,
PHE-v2-09,STA-hbt01,PHE-hbt02,PHE-hbt04,STA-hbt05z,PHE-hbt07,PHE-hbt08,ALI-spe11z,ALI-spe12z,ALI-spe13z,
ALI-spe12,ALI-spe13,ALI-v2-10,ALI-v2vn-11,ALI-v2-11z,ALI-v2v3-11,ALI-hbt11,ALI-hbt11z,BCL13,YMS13,FHL13}}.
They reflect the characteristics of the particle-emitting sources in different aspects at different stages.
A combined investigation of these observables can provide very strong constraints for the source models.
So far, much progress has been made in understanding the experimental data of the heavy ion collisions at
the top energies of the Relativistic Heavy Ion Collider (RHIC) \cite{{RHIC_exp-rep,QM06exp,QM06theor,
QM08exp,QM08sum,QM09sum,QM11exp,QM11theor,QM12sum}}.
However, more detailed investigations of the physics beneath the data through multi-observable analyses are
still needed.  On the other hand, the experimental data of several TeV heavy ion collisions at the Large
Hadron Collider (LHC) have been recently published \cite{{ALI-spe11z,ALI-spe12z,ALI-spe13z,ALI-spe12,
ALI-spe13,ALI-v2-10,ALI-v2vn-11,ALI-v2-11z,ALI-v2v3-11,ALI-hbt11,ALI-hbt11z,ALI-y-11,CMS-v2-11,
ALI-vn-12,LHC-exp12}}.  It is an ambitious goal for the models to explain consistently the data of the heavy
ion collisions at the RHIC and LHC energies.

In Refs. \cite{WNZ04,WNZ06,ZTY09,WNZ09}, W. N. Zhang {\it et al.} proposed and developed a granular source
model of QGP droplets to explain the HBT data of the RHIC experiments \cite{{STA-hbt01,PHE-hbt02,PHE-hbt04,
STA-hbt05z,PHE-hbt07,PHE-hbt08}}.  Their investigations indicate that the short evolution lifetime and wide
initial distribution of the QGP droplets in the granular source can lead to the result of the HBT radii,
$R_{\rm out} \sim R_{\rm side}$.  Here the labels ``out" and ``side" denote the transverse directions
parallel and perpendicular to the transverse momentum of the pion pair \cite{Ber88,Pra90}.  And, the granular
source results of the pion transverse momentum spectrum \cite{WNZ06,WNZ09}, elliptic flow \cite{WNZ06},
and HBT radii \cite{WNZ06,WNZ09} are well in agreement with the experimental measurements for the Au-Au
collisions at $\sqrt{s_{NN}}=$ 200 GeV at the RHIC \cite{PHE-spe04z,PHE-v2-03,PHE-hbt04,STA-hbt05z}.  They
also find \cite{ZTY09,WNZ09} that the granular source model may reproduce the main characteristics of the
two-pion source functions, extracted by the imaging techniques 
\cite{BroDan97,BroDan98,BroDan01,DanPra05,DanPra07},
in the RHIC experiments \cite{PHE-hbt07,PHE-hbt08}.  In Refs. \cite{CYW04,WNZ05}, the fluctuating signatures
of the single-event HBT correlation functions of granular sources are investigated.  The detection of source
inhomogeneity through the fluctuating single-event HBT correlation functions is discussed in Ref. \cite{YYR08},
with the smoothed particle hydrodynamics (SPH) \cite{Agu01,Ham04}.  Recently, the HBT analyses in the granular
source model for the experimental data of the most central Pb-Pb collisions at $\sqrt{s_{NN}} =$ 2.76 TeV at
the LHC \cite{ALI-hbt11} are performed \cite{WNZ11}.  The model parameters of the granular sources for the
most central collisions at the RHIC and LHC energies are compared and discussed \cite{WNZ11}.

Although the granular source model explained the pion HBT radii in the most central Au-Au and Pb-Pb collisions
at the RHIC and the LHC respectively \cite{WNZ11}, it is still a challenge to models to  explain consistently 
the experimental HBT measurements in the different centrality regions of the collisions at the RHIC and the 
LHC energies.  On the other hand, pion momentum spectrum and elliptic flow are very sensitive to collision centrality.  A combined investigation of pion momentum spectrum, elliptic flow, and HBT interferometry in the 
different centrality regions of the collisions  at the RHIC and the LHC  energies, in the granular source model, 
is of great interest.  In this work, we systematically investigate the pion transverse momentum spectrum,
elliptic flow and HBT interferometry in the granular source model for the heavy ion collisions at the
RHIC and LHC energies with different centralities.  By comparing the granular source results of pion
transverse momentum spectrum, elliptic flow, and HBT radii with the experimental data of the Au-Au
collisions at $\sqrt{s_{NN}}=$ 200 GeV with 0--5\%, 10--20\%, and 30--50\% centralities, and the experimental data of the Pb-Pb collisions at $\sqrt{s_{NN}}=$ 2.76 TeV with 10--20\% and 40--50\% centralities, we obtain
the model parameters as a function of collision centrality and energy.  We investigate the space-time
structure and expansion velocities of the granular sources at the RHIC and LHC energies with the
different centralities.  Our investigations indicate that the granular source model can reproduce the
experimental data of the pion transverse momentum spectra, elliptic flow, and HBT radii of the Au-Au
collisions at RHIC \cite{{STA-spe04z,PHE-spe04z,STA-v2-05z,STA-hbt05z}} and the Pb-Pb collisions at
LHC \cite{{ALI-spe13z,ALI-v2-11z,ALI-hbt11z}}.  The parameters in the granular source model exhibit
certain regularities for collision centrality and energy.  The space-time structure and expansion
velocities of the granular source are consistent with that reflected by the observables.

The rest of this paper is arranged as follows.  In Sec. II, we describe the basic ingredients of the 
granular source model used in this work.  In Sec. III, we present the pion transverse momentum spectrum, 
elliptic flow and HBT results of the granular sources for the heavy ion collisions at the RHIC and LHC 
energies with different centralities.  The regularities of the model parameters are also discussed in 
this section.  In Sec. IV, we investigate the space-time structure and expansion velocities of the 
granular sources.  Finally, the summary and discussions are given in Sec. V.

\section{Granular source model}

In the heavy ion collisions at the RHIC top energies and the LHC energy, the created strong-coupled QGP 
(sQGP) systems in the central rapidity region may reach local equilibrium at a very short time, and then 
expand rapidly along the beam direction ($z$-axis).  Because of the random variations in the distribution 
of collision nucleons due to quantum fluctuations, the local equilibrium system is not uniform in the 
transverse plane ($x$-$y$ plane) \cite{ISFFSC12}.  It may form many tubes along the beam direction during 
the subsequent fast longitudinal expansion and finally fragment into many QGP droplets with the effects 
of ``sausage" instability, surface tension, and bulk viscosity \cite{WNZ06,Tak09,CYW73,Tor08}.  
As a first-step idealized approximation, granular source model regards the whole source evolution as 
the superposition of the individual evolutions of the QGP droplets.  Each droplet has a position-dependent 
initial velocity and evolves hydrodynamically.  

As in Ref. \cite{WNZ06}, we suppose the QGP droplets in the granular source initially distribute within a
cylinder along $z$-axis by
\begin{eqnarray}
\frac{dN_d}{dx_0dy_0dz_0}&{\propto}&\Big[1-e^{-(x_0^2+y_0^2)/\Delta {\cal R}_T^2}\Big]\theta({\cal R}_T-
\rho_0)\cr
&&\times \theta({\cal R}_z -|z_0|),
\end{eqnarray}
where $\rho_0=\sqrt{x_0^2+y_0^2}$  and $z_0$ are the initial transverse and longitudinal coordinates
of the droplet centers.  The parameters ${\cal R}_T$ and ${\cal R}_z$ describe the initial sizes of
the source, and $\Delta {\cal R}_T$ is a shell parameter in the droplet frame \cite{WNZ06}.

In Ref. \cite{WNZ11}, the Bjorken hypothesis \cite{Bjo83} is used to describe the longitudinal velocity
of droplet for the most central collisions, and the transverse velocity of droplet has a form of
exponential power.  Considering the longitudinal velocity of droplet varying with collision centrality,
we also introduce a longitudinal power parameter, which will be determined by experimental data, to
describe the longitudinal velocity phenomenologically.  The initial velocities of the droplets in granular
source frame are assumed as \cite{WNZ06}
\begin{equation}
\label{ini_v}
v_{{\rm d}i}=\mathrm{sign}(r_{0i}) \cdot a_i \bigg(\frac{|r_{0i}|}{{\cal R}_i}\bigg)^{b_i},~~~~~~
i=1,\,2,\,3,
\end{equation}
where $r_{0i}$ is $x_0$, $y_0$, or $z_0$ for $i=$ 1, 2, or 3, and $\text{sign}(r_{0i})$ denotes the
signal of $r_{0i}$, which ensures a outward droplet velocity.  In Eq. (\ref{ini_v}), ${\cal R}_i =
({\cal R}_T, {\cal R}_T, {\cal R}_z)$, $a_i=(a_x,a_y,a_z)$ and $b_i=(b_x,b_y,b_z)$ are the magnitude
and exponent parameters in $x$, $y$, and $z$ directions, which are associated with the early
thermalization and pressure gradients of the system at the fragmentation.  It is convenient to use
the equivalent parameters $\overline{a}_T =(a_x+a_y)/2$ and $\Delta a_T=a_x-a_y$ instead of $a_x$
and $a_y$.  The parameters $\overline{a}_T$ and $\Delta a_T$ describe the transverse expansion and
asymmetric dynamical behavior of the system at the fragmentation, respectively.  For simplicity,
we take $b_x=b_y=b_T$ in calculations.  The parameters $b_T$ and $b_z$ describe the coordinate
dependence of exponential power in transverse and longitudinal directions.

In the calculations of the hydrodynamical evolution of the droplet, we use the equation of state
(EOS) of the S95p-PCE165-v0 \cite{She10}, which combines the lattice QCD data at high temperature
with the hadron resonance gas at low temperature.  We assume systems fragment when reaching a
certain local energy density, and take the initial energy density of the droplets to be 2.2
GeV/fm$^3$ for all considered collisions for simplicity \cite{WNZ11}.  The initial droplet radius
is supposed satisfying a Gaussian distribution with the standard deviation $\sigma_d=$ 2.5 fm in
the droplet local frame \cite{WNZ11}.

The final identical pions are considered to be emitted out of the surfaces of droplets with momenta
obeying the Bose-Einstein distribution in the local frame at freeze-out temperature $T_f$.  To
include the resonance decayed pions later as well as the directly produced pions at chemical freeze
out early, a wide region of $T_f$ is considered with the probability \cite{WNZ11}
\begin{eqnarray}
\frac{dP}{dT_f} &{\propto}& f_{\text{dir}}\,e^{-\frac{T_{\text{chem}}-T_f}{\Delta T_{\text{dir}}}}
+({1-f_{\text{dir}}})\cr
&\times& e^{-\frac{T_{\text{chem}}-T_f}{\Delta T_{\text{dec}}}}, ~~(T_{\text{chem}}>T_f>80~\text{MeV}),
\end{eqnarray}
where $f_{\text{dir}}$ is the fraction of the direct emission around the chemical freeze out temperature
$T_{\text{chem}}$,  $\Delta T_{\text{dir}}$ and $\Delta T_{\text{dec}}$ are the temperature widths for the
direct and decay emissions, respectively.  In the calculations, we take $f_{\text{dir}}=0.75$, $\Delta
T_{\text{dir}}=10$ MeV, and $\Delta T_{\text{dec}}=90$ MeV as in Ref. \cite{WNZ11}.  The value of
$T_{\text{chem}}$ is taken to be 165 MeV as it be taken in the S95p-PCE165-v0 EOS \cite{She10}.

After fixing the parameters used in the calculations of hydrodynamical evolution and freeze-out temperature,
the free model parameters are the three source geometry parameters ($R_T$, $\Delta R_T$, $R_z$) and the five
droplet velocity parameters ($\overline{a}_T$, $\Delta a_T$, $a_z$, $b_T$, $b_z$).  They are associated with
the initial size, expansion, and directional asymmetry of system, and have significant influence on the
observables of pion momentum spectra, elliptic flow, and HBT radii.  We will combine the experimental data
of these observables to investigate the parameters of the granular source as a function of the collision
centrality and energy in the heavy ion collisions at the RHIC and LHC energies in next sections.

\section{Pion momentum spectrum, elliptic flow, and HBT results}

In high energy heavy ion collisions, the invariant momentum distribution of final particles can be written
in the form of a Fourier series \cite{SV-YZ96,AP-SV98},
\begin{eqnarray}
\label{pdis}
E\frac{d^3N}{d^3p}=\frac{1}{2\pi}\frac{d^2N}{p_Tdp_Tdy}\left[1+\sum_n{2v_n\cos(n\phi)}\right],
\end{eqnarray}
where $E$ is the energy of the particle, $p_T$ is the transverse momentum, $y$ is the rapidity (it should
not bring a mistake with coordinate from the context), and $\phi$ is the azimuthal angle with respect to
the reaction plane.  In Eq. (\ref{pdis}), the first term on right is the transverse momentum spectrum in
the rapidity region $dy$, and the second harmonic coefficient $v_2$ in the summation is called elliptic
flow.

At the RHIC and LHC energies, the spectators in nucleus-nucleus collisions depart from the reaction region
rapidly after collision, and a very hot and dense fireball is formed in an almond shape perpendicular to
the reaction plane.  The particle spectra of transverse momentum at low $p_T$ ($p_T\,\undersim<\,3$ GeV/$c$)
contain the information about the transverse expansion and thermal properties of the particle-emitting
sources at freeze-out temperature \cite{{ALI-spe11z,ALI-spe12z,ALI-spe13z,ALI-spe12,ALI-spe13}}.  By
comparing the pion transverse momentum spectra of the granular sources with experimental data, we can
constrain the velocity parameters $\overline{a}_T$ and $a_z$ of the granular sources.

Choosing $x$ axis on the reaction plane, elliptic flow $v_2$ can be expressed as
\begin{equation}
v_2(p_T)=\left\langle{\cos(2\phi)}\right\rangle=\left\langle{\frac{p_x^2-p_y^2}{p_T^2}}\right\rangle.
\end{equation}
Since the reaction plane orientation is hardly to estimate exactly in experiment, an alternative technique
for elliptic flow analysis is the measurement of the two-particle cumulant of azimuthal correlations,
$v_2^2\{2\}$ \cite{{Bor01,STA-v2-02,PHE-v2-02,STA-v2-05z,PHE-v2-09,ALI-v2-10,ALI-v2vn-11,ALI-v2-11z,
ALI-v2v3-11}}, which gives essentially the same results as the reaction-plane method
\cite{STA-v2-02,STA-v2-05z,PHE-v2-09,BhaOll06}.

In non-central nucleus-nucleus collisions, the initial space-asymmetry of the system can bring anisotropic
pressure gradients, which lead to the anisotropic transverse-momentum distributions of final particles and
nonzero $v_2$.  For the granular source, the results of elliptic flow are very sensitive to the parameters
$\Delta a_T$, $b_T$, and $b_z$ \cite{WNZ06}.  The experimental data of the transverse momentum spectrum and
elliptic flow impose strict constraints on the velocity parameters in the granular source model.

Two-particle Hanbury-Brown-Twiss(HBT) correlation function is defined as the ratio of the two-particle
momentum spectrum $P(\bp_1,\bp_2)$ to the product of two single-particle momentum spectra $P(\bp_1)P(\bp_2)$.
It has been widely used to extract the space-time geometry, dynamic and coherence information of the
particle-emitting source in high energy heavy ion collisions \cite{Gyu79,Wongbook,Wie99,Wei00,Lisa05}.
In the usual HBT analysis in high energy heavy ion collisions, the two-pion correlation functions are fitted
by the Gaussian parameterized formula
\begin{equation}
\label{CF}
C(q_{\rm out},q_{\rm side},q_{\rm long})\!=\!1 \!+ \lambda\, e^{-R_{\rm out}^2 q_{\rm out}^2
-R_{\rm side}^2 q_{\rm side}^2 -R_{\rm long}^2 q_{\rm long}^2},
\end{equation}
where $q_{\rm out}$, $q_{\rm side}$, and $q_{\rm long}$ are the Bertsch-Pratt variables \cite{Ber88,Pra90},
which denote the components of the relative momentum $\bq=\bp_1-\bp_2$ in transverse out and side directions
and in longitudinal direction, respectively.  In Eq. (\ref{CF}) $\lambda$ is chaotic parameter of source,
$R_{\rm out}$, $R_{\rm side}$, and $R_{\rm long}$ are the HBT radii in out, side, and long directions.
The results of HBT radii are related to the source geometry as well as expansion.  We can finally determine
the geometry parameters ($R_T$, $\Delta R_T$, $R_z$) and the velocity parameters ($\overline{a}_T$, $\Delta
a_T$, $a_z$, $b_T$, $b_z$) in the granular source model by the multi-observable analyses of the pion spectrum,
elliptic flow, and HBT interferometry.

In the multi-observable analyses, we choose the experimental data of the Au-Au collisions at $\sqrt{s_{NN}}
=200$ GeV at the RHIC \cite{STA-spe04z,PHE-spe04z,STA-v2-05z,STA-hbt05z} and the Pb-Pb collisions at
$\sqrt{s_{NN}}=2.76$ TeV at the LHC \cite{ALI-spe13z,ALI-v2-11z,ALI-hbt11z} for determining the
geometry and velocity parameters of the granular sources.  These experimental data provide the identical
pion $p_T$ spectrum, elliptic flow, and HBT radii simultaneously in the same centrality regions (RHIC:
0--5\%, 10--20\% and 30--50\%; LHC: 10--20\% and 40--50\%).

\begin{figure*}
\includegraphics[angle=0,scale=0.55]{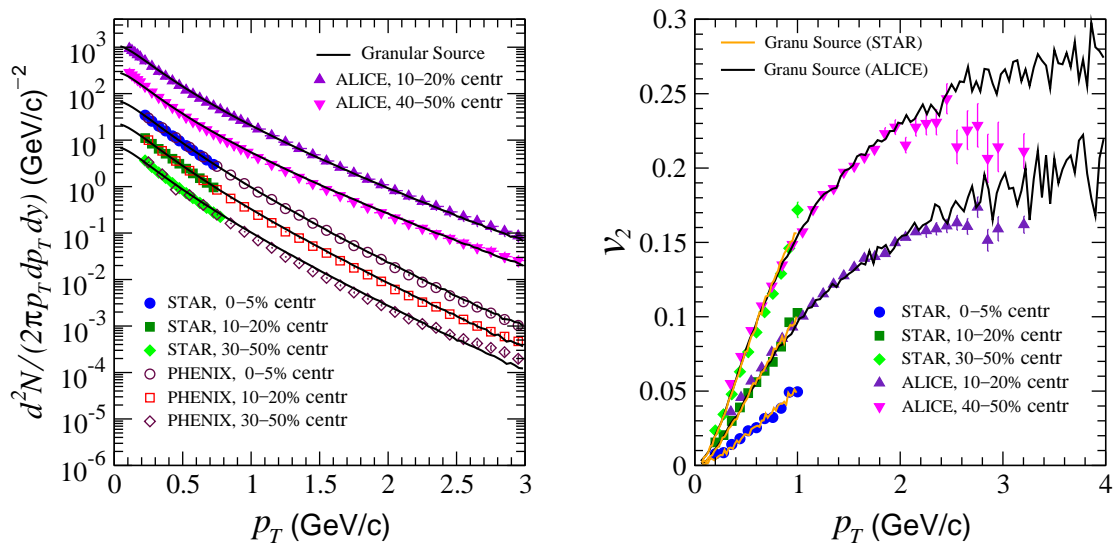}
\caption{(Color online) Left panel: the pion transverse momentum spectra of the granular sources (solid
lines) and the experimental data of negative pions in the Au-Au collisions at $\sqrt{s_{NN}}=$ 200 GeV
\cite{STA-spe04z,PHE-spe04z} and the Pb-Pb collisions at $\sqrt{s_{NN}}=2.76$ TeV \cite{{ALI-spe13z}}.
Right panel: the elliptic flow of the granular source (solid lines) and the experimental data of negative
pions in the Au-Au collisions at $\sqrt{s_{NN}}=$ 200 GeV \cite{STA-v2-05z} and the Pb-Pb collisions at
$\sqrt{s_{NN}}=2.76$ TeV \cite{ALI-v2-11z}. }
\label{fspev2}
\end{figure*}

In Fig. \ref{fspev2}, we plot the pion transverse momentum spectra (left panel) and elliptic flow (right
panel) of the granular sources for the heavy ion collisions at the RHIC and LHC energies with different
centralities.  The experimental data of the pion transverse momentum spectra and elliptic flow of the
Au-Au collisions with 0--5\%, 10--20\% and 30--50\% centralities at $\sqrt{s_{NN}}=200$ GeV at the RHIC
\cite{STA-spe04z,PHE-spe04z,STA-v2-05z}, and the Pb-Pb collisions with 10--20\% and 40--50\% centralities
at $\sqrt{s_{NN}}=2.76$ TeV at the LHC \cite{{ALI-spe13z,ALI-v2-11z}} are shown simultaneously.  Comply
with the experimental measurements, we use the rapidity cuts $|\,y\,| <0.1$ \cite{STA-spe04z} and $|\,y\,|
<0.5$ \cite{ALI-spe12z,ALI-spe13z} in the calculations of the $p_T$ spectra of the granular sources at the
RHIC energy and the LHC energy, respectively.  In the calculations of the elliptic flow of the granular
sources, the pseudorapidity cuts $|\,\eta\,|<1.0$ and $|\,\eta\,|<0.8$ are adopted as the same as in the
experimental analyses at the RHIC \cite{STA-v2-05z} and LHC \cite{ALI-v2-11z}, respectively.

As shown in Fig. \ref{fspev2}, the pion transverse momentum spectra of the granular sources agree with
the experimental data with different centralities at the RHIC and LHC energies simultaneously.  The
spectra at the LHC energy exhibit clear up-warp at $p_T>1.5$ GeV/c as compared to those at the RHIC
energy.  However, the results of elliptic flow at the LHC energy almost match the elliptic flow results
at the RHIC energy with the same and near centrality regions.  The results of elliptic flow of granular
sources exhibit clear centrality dependence as the experimental data with.  The $v_2$ results decrease
with increasing collision centrality.  At $p_T>2.5$ GeV/$c$, the granular source results of elliptic
flow are a little higher than those of experimental data.  It reflects the limitations at high $p_T$ of
the model based on hydrodynamical evolution.

\begin{figure}[!htbp]
\includegraphics[scale=0.60]{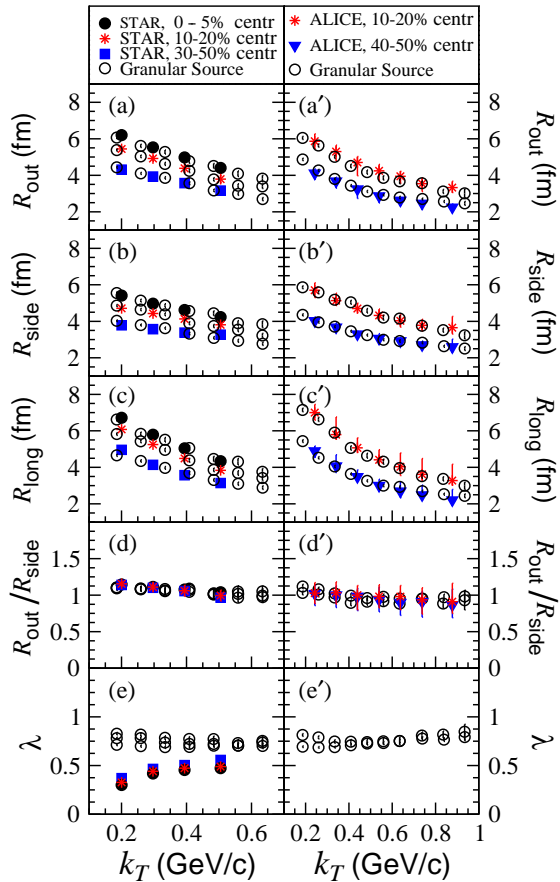}
\vspace*{2mm}
\caption{(Color online) The results of pion HBT radii and chaotic parameter of the granular sources (open
circle) with different centralities for the highest RHIC energy (left column) and LHC energy (right
column).  The solid circle, star, and solid diamond symbols in the left column are the STAR data of the
Au-Au collisions at $\sqrt{s_{NN}}=200$ GeV with 0--5\%, 10--20\% and 30--50\% centralities \cite{STA-hbt05z},
respectively.  The star and solid triangle-down symbols in the right column are the ALICE data of the Pb-Pb
collisions at $\sqrt{s_{NN}}=2.76$ TeV with 10--20\% and 40--50\% centralities \cite{ALI-hbt11z}, respectively.}
\label{fhbtfit}
\end{figure}

In Fig. \ref{fhbtfit}, we show the granular source results of the pion HBT radii and chaotic parameter as a
function of the transverse momentum of pion pair, $k_T=|\bp_1+\bp_2|/2$, obtained by Gaussian parameterized
formula fit in the longitudinally comoving system (LCMS) \cite{Wie99,Lisa05}.  The experimental data of
STAR Au-Au collisions at $\sqrt{s_{NN}}=200$ GeV \cite{STA-hbt05z} and ALICE Pb-Pb collisions at
$\sqrt{s_{NN}}=2.76$ TeV \cite{ALI-hbt11z}, which have the same centralities as the experimental data of
pion $p_T$ spectra and $v_2$ shown in Fig. \ref{fspev2}, are also plotted for comparing.  In the HBT analyses
of the granular sources for the RHIC and LHC energies, we applied the rapidity cut $|\,y\,|<0.5$ and the
pseudorapidity cut $|\,\eta\,|<0.8$ as the same in the experimental analyses of STAR \cite{STA-hbt05z} and
ALICE \cite{ALI-hbt11z}, respectively.

From Fig. \ref{fhbtfit} it can be seen that the HBT radii $R_{\rm out}$, $R_{\rm side}$, and $R_{\rm long}$
of the granular sources simultaneously agree with the experimental data at the RHIC and LHC energies.  Both
the transverse and longitudinal HBT radii increase with increasing collision centrality.  At the LHC energy 
the results of the HBT radii $R_{\rm out}$, $R_{\rm side}$, and $R_{\rm long}$ are larger than those at the 
RHIC energy, respectively.  However, the results of the ratio of $R_{\rm out}/R_{\rm side}$ are always about 
1 and independent of the collision centrality and energy.  In experimental HBT measurements, many effects, 
such as the Coulomb interaction between the final particles, particle missing-identification, source 
coherence, {\it etc.}, can influence the results of the chaotic parameter $\lambda$ 
\cite{Gyu79,Wongbook,Wie99,Wei00,Lisa05}.
Because these effects do not be considered in the granular source model, the $\lambda$ results of the 
granular sources are larger than the experimental data.

\begin{table*}
\begin{center}
\caption{The geometry and velocity parameters of the granular sources. }
\begin{tabular}{l|ccccccccc}
\hline\hline
~~~Centrality&~${\cal R}_T\,\text{(fm)}$~&~$\Delta {\cal R}_T\,\text{(fm)}~$&~${\cal R}_z\,\text{(fm)}$~&
~~~~~$\overline{a}_T$~~~~~&~~~~~$\Delta a_T$~~~~~&~~~~~$a_z$~~~~~&~~~~~$b_T$~~~~~&~~~~~$b_z$~~~~~\\
\hline
~RHIC,~~~0--5~\% & 5.8 & 0.7 & 3.9 & 0.469 & 0.066 & 0.593 & 0.76 & 0.13 \\
~RHIC,~10--20\%  & 4.5 & 0.5 & 2.9 & 0.454 & 0.115 & 0.593 & 0.56 & 0.11 \\
~RHIC,~30--50\%  & 2.5 & 0.3 & 0.5 & 0.437 & 0.156 & 0.593 & 0.37 & 0.06 \\
\hline
~\,LHC,\,~10--20\% & 6.0 & 0.9 & 5.5 & 0.431 & 0.092 & 0.592 & 0.35 & 0.13 \\
~\,LHC,\,~40--50\% & 2.5 & 0.4 & 1.8 & 0.407 & 0.131 & 0.590 & 0.23 & 0.03 \\
\hline\hline
\end{tabular}
\end{center}
\end{table*}

In Table I we present the values of the geometry and velocity parameters of the granular sources used
in the multi-observable analyses.  One can see that the value of ${\cal R}_T$ for a certain collision
centrality and energy is larger than that of ${\cal R}_z$.  So the initial geometry of the granular
source is a short cylinder.  The initial transverse and longitudinal sizes of the granular sources
${\cal R}_T$ and ${\cal R}_z$ increase with increasing collision centrality.  The transverse shell 
parameter $\Delta {\cal R}_T$ also increases with increasing collision centrality.  Because $\Delta 
{\cal R}_T\ll{\cal R}_T$, the shell effect is small and the initial distributions of droplets are 
almost volume distribution.  
For 10--20\% centrality, the values of the geometry parameters ${\cal R}_T$ and ${\cal R}_z$ of the
granular source for the RHIC energy are smaller than that for the LHC energy, respectively.  The QGP
droplets in the granular sources initially distribute in larger transverse and longitudinal regions
for the more central and higher energy collisions.

\begin{figure}[!htbp]
\vspace*{3mm}
\includegraphics[scale=0.58]{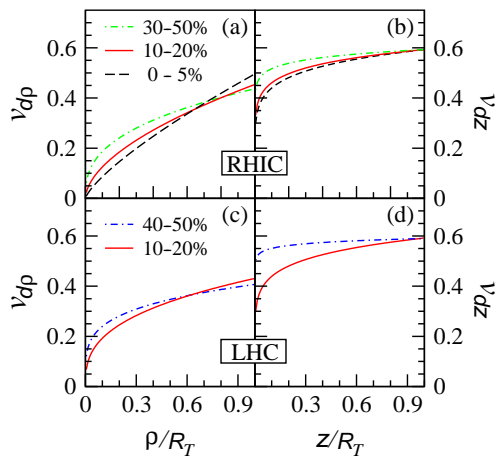}
\vspace*{1mm}
\caption{(Color online) The droplet transverse and longitudinal velocities $v_{d\rho}=\overline{a}_T
(\rho/{\cal R}_T)^{b_T}$ and $v_{dz}=a_z (|z|/{\cal R}_z)^{b_z}$.}
\label{fvdrz}
\end{figure}

In transverse direction, the velocity parameters $\overline{a}_T$ and $b_T$ increases with increasing 
collision centrality.  For a fixed $b_T$, the larger the parameter ${\overline a}_T$, the larger the 
average transverse velocity of droplet is.  Because the values of $b_T$ are less than one, the larger 
$b_T$, the smaller the average transverse velocity of droplet is, if $\overline{a}_T$ is fixed.
In longitudinal direction, the parameter $a_z$ is almost independent of collision centrality and
energy.  The values of the parameter $b_z$ are much smaller than those of $b_T$, while $b_z$
increases with increasing collision centrality as $b_T$.  The large difference between the values of 
the transverse and longitudinal exponent parameters $b_T$ and $b_z$, and the different centrality
dependence of $\overline{a}_T$ and $a_z$ reflect the different dynamical behaviors in transverse
and longitudinal directions in the heavy ion collisions at the RHIC and LHC energies.  In Fig.
\ref{fvdrz} we plot the droplet velocities $v_{d\rho}=\overline{a}_T (\rho/{\cal R}_T)^{b_T}$ and
$v_{dz}=a_z(|z|/{\cal R}_z)^{b_z}$ of the granular sources.  One can see that the droplet transverse
velocity of central collision is smaller than that of peripheral collision in the center region of
the source, although the droplet transverse velocity of central collision is larger at the edge of
the source.  The average longitudinal velocity of droplet is larger than the average transverse
velocity.  In peripheral collisions, the droplet longitudinal velocity is almost a constant in
the source.  From Table I, it can be seen that the velocity parameter $\Delta a_T$ decreases with
collision centrality.  This leads to the increase of $v_2$ with decreasing collision centrality.

\section{Granular source space-time and expansion}

In the granular source model, the results of single particle $p_T$ spectrum, elliptic flow, and HBT
radii are strongly related to the source space-time and expansion properties.  The successes of the
granular source model in explaining the experimental data of pion $p_T$ spectrum, elliptic flow, and
HBT radii in the heavy ion collisions at the RHIC and LHC inspire us to further study the granular
source space-time and expansion features.

\begin{figure}[!htbp]
\begin{center}
\begin{minipage}{0.18\textwidth}
\hspace*{-12mm}\includegraphics[scale=0.45]{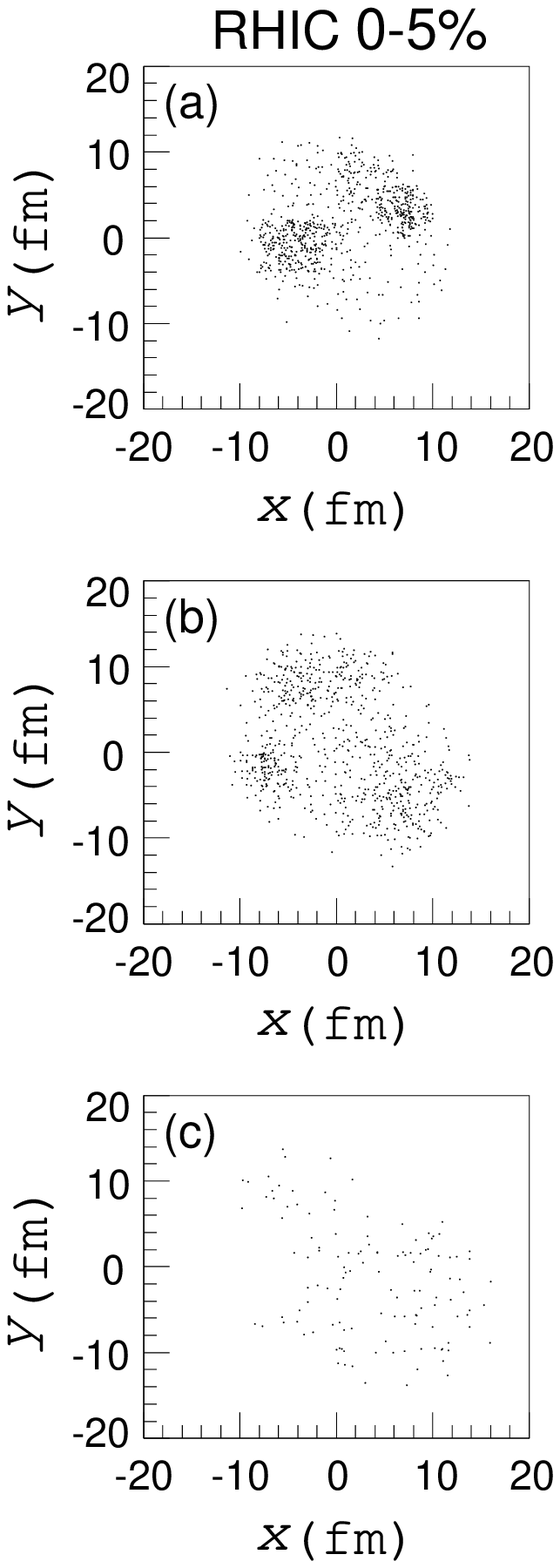}
\end{minipage}%
\begin{minipage}{0.18\textwidth}
\hspace*{-10mm}\includegraphics[scale=0.45]{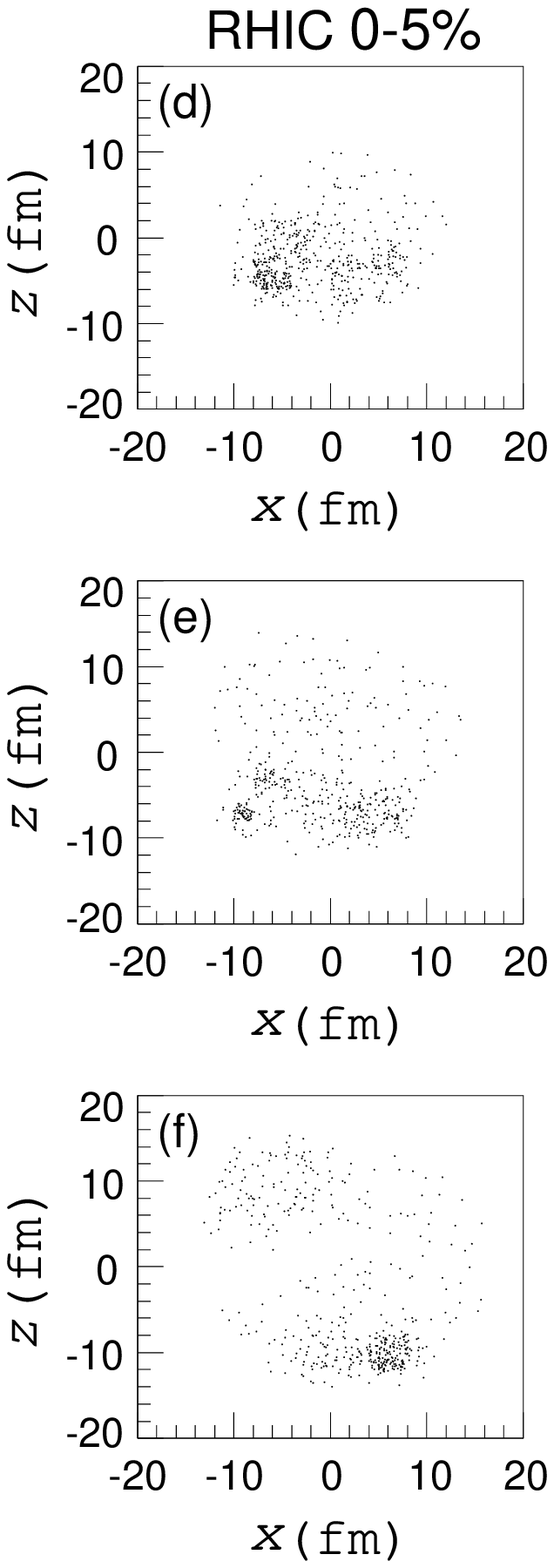}
\end{minipage}%
\caption{The pictures of the distributions of pion source points in one granular source event for the
RHIC central collision.  The $z$ region for the panels (a), (b), and (c) is $|z|<1$ fm.  The $y$ region
for the panels (d), (e), and (f) is $|y|<1$ fm.  The time for the panels (a) and (d), (b) and (e), and
(c) and (f) is $t=$ 4, 8, and 12 fm/$c$, respectively.  The exposure time for these pictures is 0.5
fm/$c$. }
\label{fdisxyz}
\vspace*{-5mm}
\end{center}
\end{figure}

In Fig.\ref{fdisxyz}, we show the pictures of the distributions of pion source points in one granular
source event with $4\times 10^5$ pion pairs for the heavy ion collisions at the RHIC energy with 0--5\%
centrality.  The $z$ region for the $x$-$y$ distributions [panels (a), (b), and (c)] is $|z|<1$ fm.  The
$y$ region for the $x$-$z$ distributions [panels (d), (e), and (f)] is $|y|<1$ fm.  The time for the panels
(a) and (d), (b) and (e), and (c) and (f) is $t=$ 4, 8, and 12 fm/$c$, respectively.  For each picture
the exposure time is 0.5 fm/$c$.  One can see that the distributions are inhomogeneous.  There are
separated ``clumps", which correspond to the separated droplets, in the distributions.  The clump-structure
exists in the whole duration of source evolution.  In two-pion HBT analysis, the correlated two pions are
taken from the same event.  The effect of the clump-structure on the HBT results will exist even after
many-event-mixing, which smooths out the clump-structure and leads to a continued distribution of the
source points.  Because many droplets evolve simultaneously in the granular source, the source lifetime
is smaller as compared to that of a continued big source, which evolves in whole and freeze out from the
source surface \cite{WNZ04,WNZ06,WNZ07,WNZ11,WNZ11J}.  It will be seen that the distributions of
source points for many events are more volume distributions rather than surface distributions because of
the contributions from the droplets in the central region of the granular source.

\begin{figure}[!htbp]
\begin{center}
\begin{minipage}{0.18\textwidth}
\hspace*{-12mm}\includegraphics[scale=0.45]{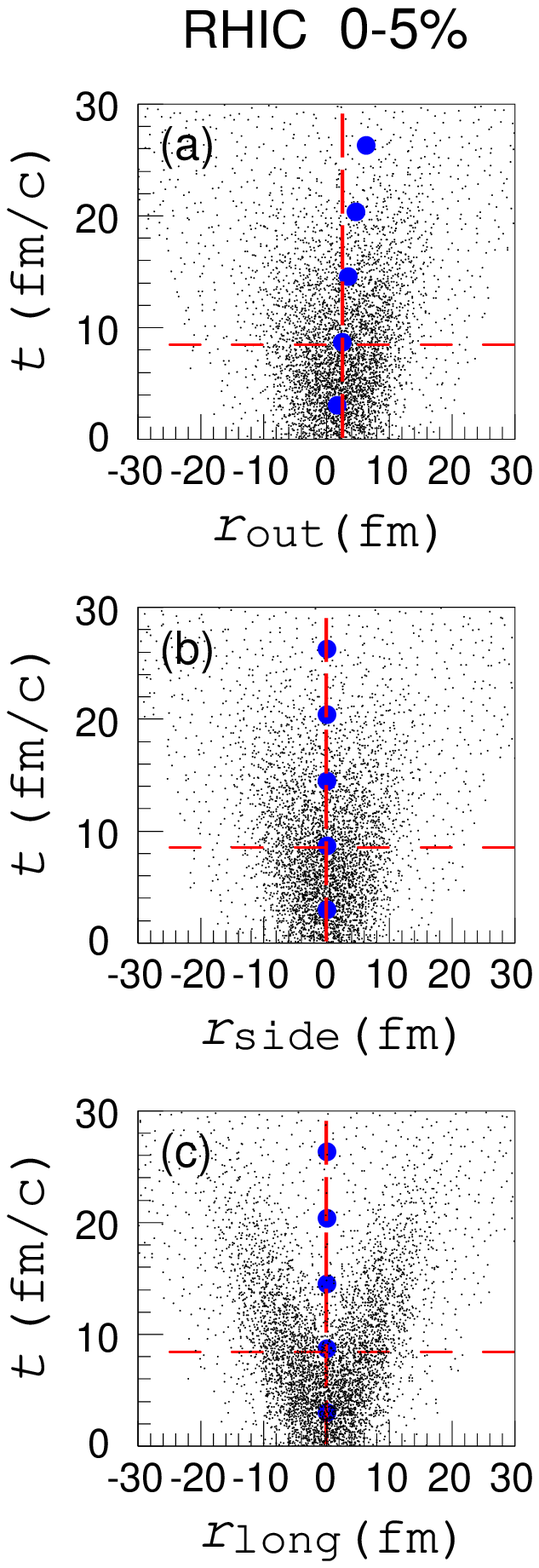}
\end{minipage}%
\begin{minipage}{0.18\textwidth}
\hspace*{-10mm}\includegraphics[scale=0.45]{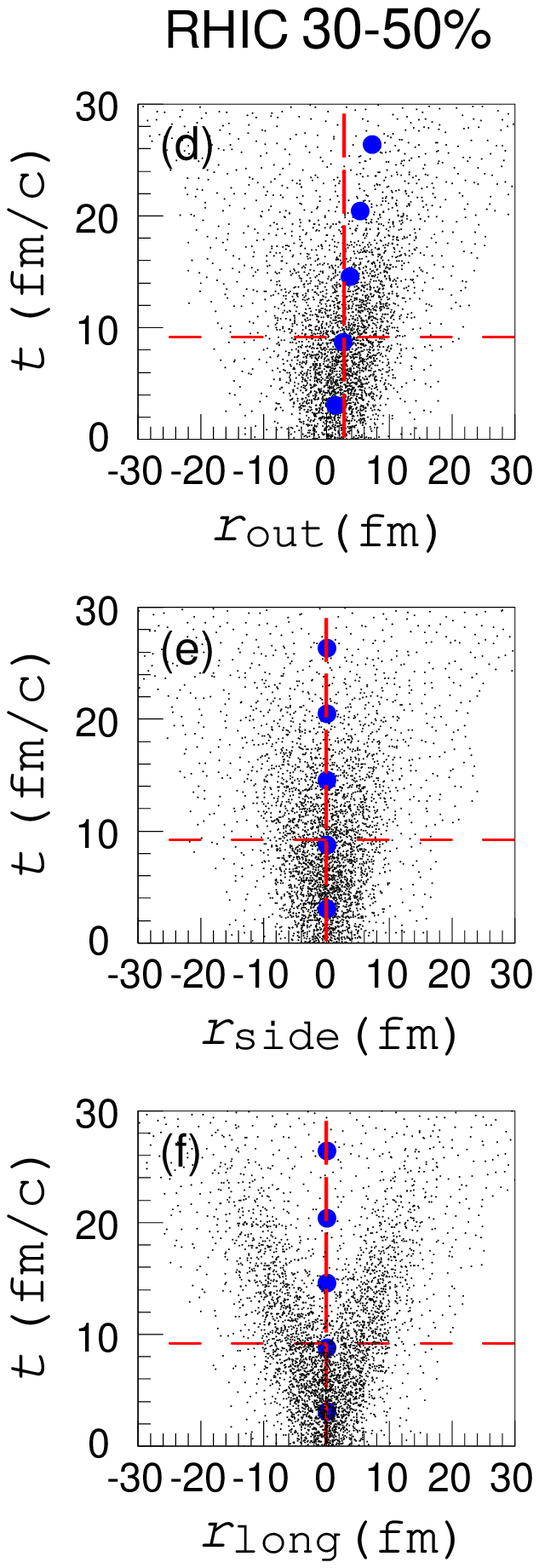}
\end{minipage}%
\caption{(Color online) The source-point distributions of final identical pions in $r_i-t$ plane
($i=\rm{out, side, long}$), for the granular sources of the central and peripheral collisions at
the RHIC energy.  The dashed lines are the average values of $r_i$ and $t$ for all of the source
points.  The bullets are the average values of $r_i$ obtained from the same $t$ bins. }
\label{ftro-disRHIC}
\vspace*{-5mm}
\end{center}
\end{figure}

\begin{figure}[!htbp]
\vspace*{3mm}
\includegraphics[scale=0.6]{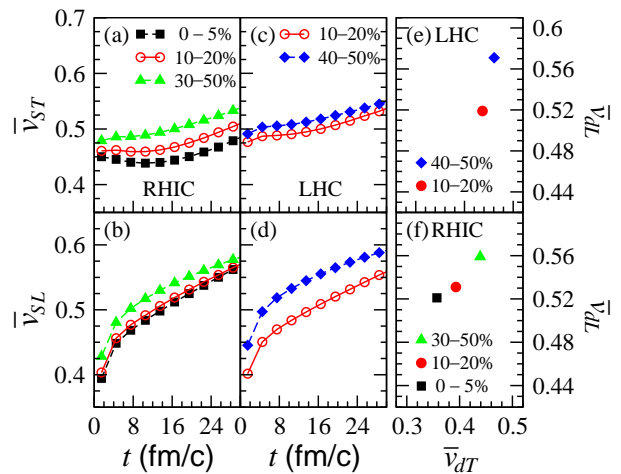}
\vspace*{2mm}
\caption{(Color online) (a)--(d) The average transverse and longitudinal velocities of the pion source
points, ${\overline v}_{ST}$ and ${\overline v}_{SL}$, versus the emission time of the granular sources.
(e) and (f) The average transverse and longitudinal droplet velocities ${\overline v}_{dT}$ and
${\overline v}_{dL}$ of the granular sources for the RHIC and LHC collisions. }
\label{F-vt}
\end{figure}

In Fig.\ref{ftro-disRHIC}, we plot the space-time distributions of the pion source points of the
granular sources projected on $t$-$r_{\rm out}$, $t$-$r_{\rm side}$, and $t$-$r_{\rm long}$ planes,
for five hundred events for the heavy ion collisions at the RHIC energy with 0--5\% and 30--50\%
centralities.  Here, the dashed lines are the average values of $r_i$ ($i=\rm{out, side, long}$)
and $t$ for all of the source points, and the bullets are the average values of $r_i$ obtained by
averaging over the same $t$ bins.  In the calculations the same rapidity cut $|\,y\,|<0.5$ as in
the experimental HBT analyses \cite{STA-hbt05z} is used.  We take the relative momentum cut $|\,
q_i\,|<100$ MeV/$c$ for pion pairs because most of the contributions in HBT correlation functions
come from the particle pairs with small relative momenta \cite{{Gyu79,Wongbook,Wie99,Wei00,Lisa05}}.
One can see that the distributions for many events are smoothed.  The widths of the $r_i$-distributions
for the central collision are wider than those for the peripheral collision.  In side and long directions,
the distributions are symmetric with respect to $r_i=0$.  However, one can observe a time-increased
asymmetry for the distributions in out direction.  It is because the coordinate-dependent source
transverse expansion boosts particle momenta along the direction out of the source, and this
coordinate-momentum correlation leads the result that the average angle between particle momentum
and emitting coordinate trends to be smaller than isotropic emission.  For the central and peripheral
collisions, the average values of $r_{\rm out}$ are 2.52 and 2.78 fm respectively.  The effect of
the coordinate-momentum correlation is larger for the peripheral collision because of the larger source
transverse velocity that the peripheral collision with (will be seen in Fig. \ref{F-vt} (a)).

\begin{figure}[!htbp]
\begin{center}
\begin{minipage}{0.18\textwidth}
\hspace*{-12mm}\includegraphics[scale=0.45]{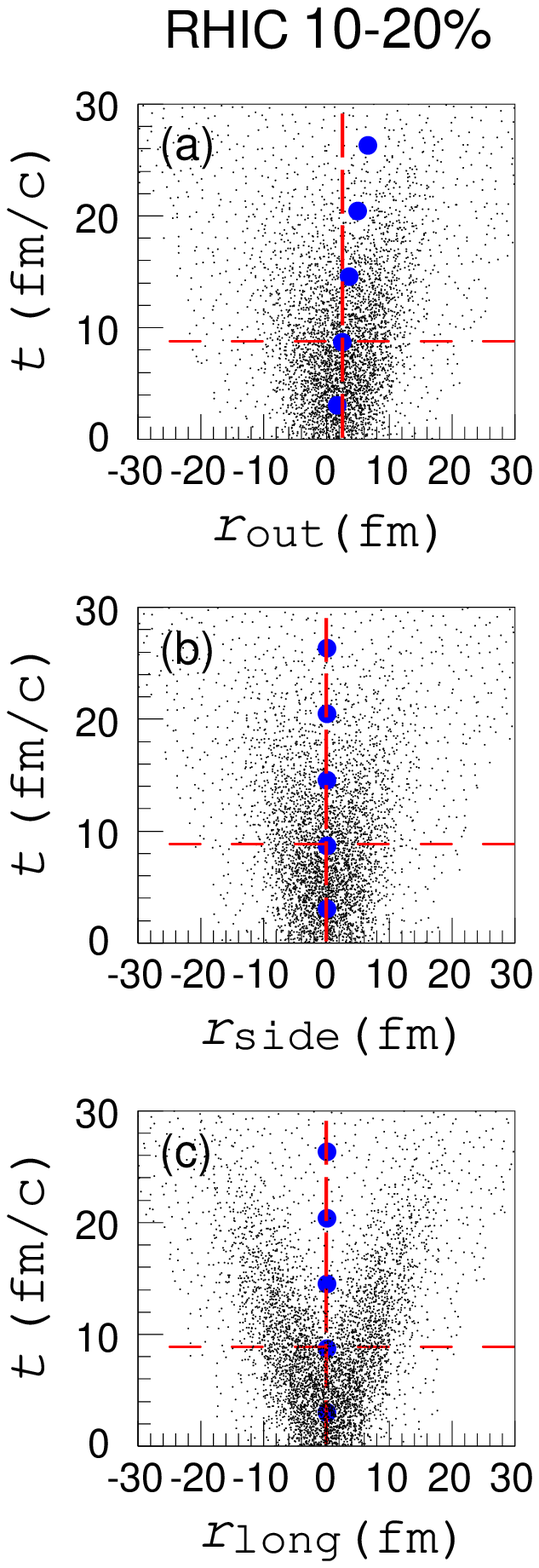}
\end{minipage}%
\begin{minipage}{0.18\textwidth}
\hspace*{-10mm}\includegraphics[scale=0.45]{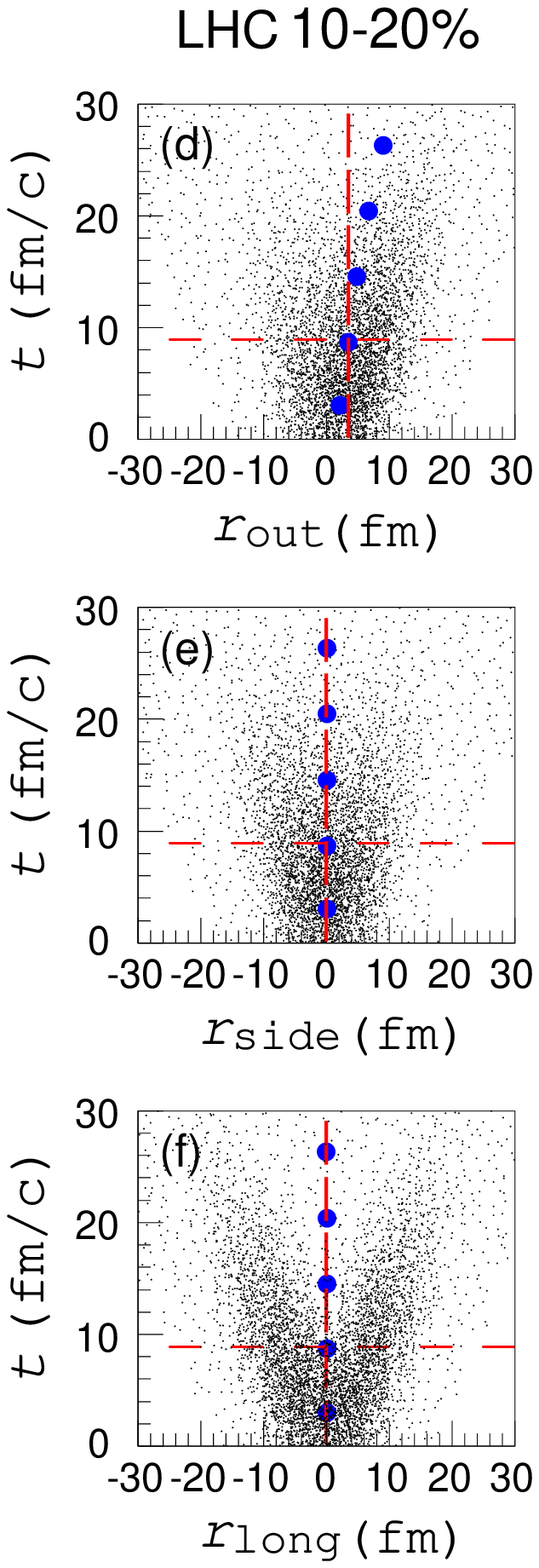}
\end{minipage}%
\caption{(Color online) The source-point distributions of final identical pions in $r_i-t$ plane
($i=\rm{out, side, long}$), for the granular sources of the RHIC and LHC collisions with 10--20\%
centrality.  The dashed lines are the average values of $r_i$ and $t$ for all of the source points.
The bullets are the average values of $r_i$ obtained from the same $t$ bins. }
\label{ftro-disC1020}
\vspace*{-5mm}
\end{center}
\end{figure}

In Fig. \ref{ftro-disC1020} we compare the space-time distributions of the pion source points of the
granular sources for the collisions at the RHIC and LHC energies with the same centrality, 10--20\%.
As in Fig. \ref{ftro-disRHIC} the dashed lines are for the average values of $r_i$ and $t$ for all
of the source points, and the bullets are the average values of $r_i$ obtained from the same $t$
bins.  We take the relative momentum cut $|\,q_i\,|<100$ MeV/$c$ in the calculations.  The rapidity
and pseudorapidity cuts $|\,y\,|<0.5$ and $|\,\eta\,|<0.8$ are taken for the RHIC and LHC collisions
respectively as in the experimental HBT analyses \cite{STA-hbt05z,ALI-hbt11z}.  It can be seen that
the widths of the $r_{\rm out}$, $r_{\rm side}$, and $r_{\rm long}$ distributions for the LHC source
are wider as compared to those for the RHIC source with the same collision centrality.  In side and
long directions, the distributions are symmetric with respect to $r_i=0$.  But the distributions in
out direction are asymmetric with respect to $r_{\rm out}=0$ because of the coordinate-momentum
correlations arising from the source transverse expansion.  The asymmetry effect is larger for the
granular source at the LHC energy ($\langle r_{\rm out}\rangle=3.59$ fm) than that at the RHIC
energy ($\langle r_{\rm out}\rangle=2.64$ fm) because of the larger transverse velocity that the
granular source with for the LHC collision than the RHIC collision.

Figure \ref{F-vt} (a) and (b) display the average transverse and longitudinal velocities of the granular
sources as a function of time at the RHIC energy.  Figure \ref{F-vt} (c) and (d) display the average
transverse and longitudinal velocities of the granular sources as a function of time at the LHC energy.
It can be seen that the average velocities increase with decreasing collision centrality at both the
energies.  The larger average transverse velocities of sources at the LHC energy are consistent with
the results that the $p_T$ spectra at the LHC energy exhibit up-warp at larger $p_T$ as compared to
those at the RHIC energy (see the left panel of Fig. \ref{fspev2}).
The differences of the transverse velocities for different centralities become small at the
higher energy, and the differences of the longitudinal velocities for different centralities are larger
at the LHC energy.  In Fig. \ref{F-vt} (e) and (f) we present the average transverse and longitudinal
droplet velocities for the granular sources at the RHIC and LHC energies.  The average transverse and
longitudinal droplet velocities increase with the decreasing collision centrality for both the energies.
In our granular source model, the droplet evolution in the local frame is independent of the collision
energy and centrality.  However, the different droplet velocities lead to the difference between the
average emission time in the source center-of-mass frame because of the different Lorentz time delays.
The larger the droplet velocity the larger the emission time is.  For example, the average emission
time of the granular sources for the RHIC central and peripheral collisions are 8.39 and 9.11 fm/$c$,
respectively (see Fig. \ref{ftro-disRHIC}).  The average emission time for the granular source for LHC
peripheral collision is the largest because the average droplet velocity is the largest in this case
(see Fig. \ref{F-vt} (e)).

In the two-pion interferometry in high energy heavy ion collisions, the difference between the transverse
HBT radii $R_{\rm out}$ and $R_{\rm side}$ satisfies \cite{Her95,Cha95,Wie99},
\begin{eqnarray}
\label{Eq-hbtRos}
R_{\rm out}^2(k_T)-R_{\rm side}^2(k_T) \approx \big[\,\langle {\beta_T\,\tilde t}^2\rangle
- 2 \langle {\beta_T\,\tilde r}_{\rm out}{\tilde t}\,\rangle \big](k_T),
\end{eqnarray}
where $\beta_T=|\bp_{1T}+\bp_{2T}|/(E_1+E_2)$ is the transverse velocity of the pair, ${\tilde t}=t-
\langle\,t\,\rangle$ and ${\tilde r}_{\rm out} = r_{\rm out} - \langle\,r_{\rm out}\,\rangle$ are the
deviations of source time and coordinate $r_{\rm out}$ from their averages, respectively.

\begin{figure}[!htbp]
\vspace*{3mm}
\includegraphics[scale=0.55]{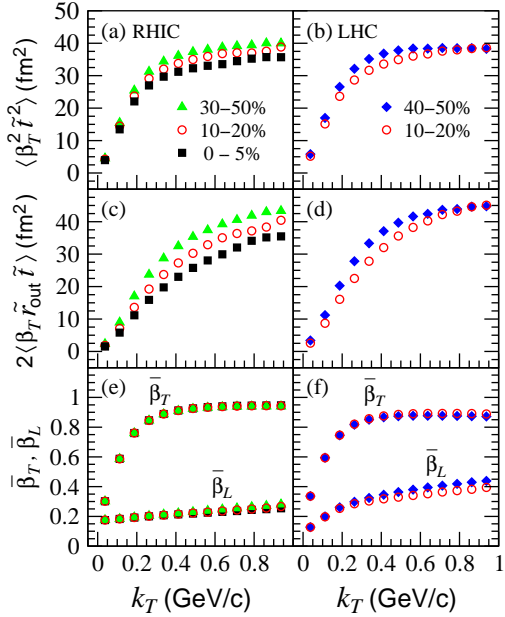}
\vspace*{1mm}
\caption{(Color online) The granular source $\langle \beta_T^2{\tilde t}^2\rangle$, $2\langle \beta_T
{\tilde r}_{\rm out} {\tilde t}\,\rangle$, ${\overline \beta}_T$, and ${\overline \beta}_L$ versus $k_T$
for the RHIC and LHC collisions. }
\label{F-rort}
\end{figure}

\begin{figure}[!htbp]
\vspace*{2mm}
\includegraphics[scale=0.50]{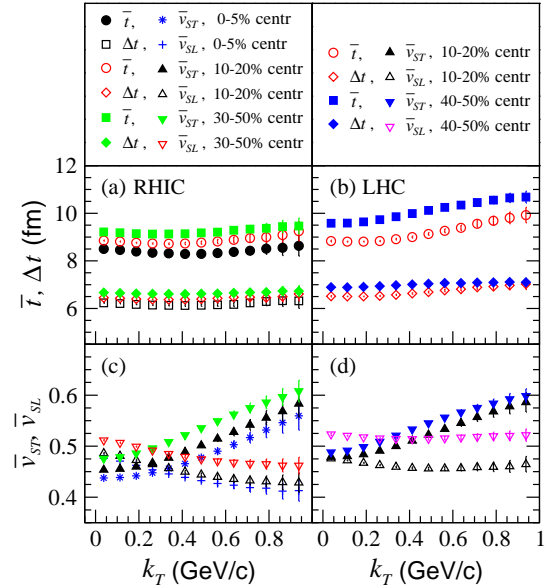}
\vspace*{1mm}
\caption{(Color online) The $k_T$ dependence of the average emission time ${\overline t}=\langle\,t\,
\rangle$, time variance root $\Delta t=[\langle {\tilde t}^2\rangle]^{1/2}$, and average source transverse
and longitudinal velocities ${\overline v}_{_{ST}}$ and ${\overline v}_{_{SL}}$ for the granular sources
at the RHIC and LHC energies. }
\label{F-tsv}
\end{figure}

In Fig. \ref{F-rort}, we plot $\langle \beta_T^2 {\tilde t}^2 \rangle$, $2\langle \beta_T {\tilde r}_{\rm
out} {\tilde t}\,\rangle$, and the average transverse and longitudinal velocities of the pion pair as a
function of $k_T$.  It can be seen that $\langle\beta_T^2{\tilde t}^2\rangle$ and $2\langle \beta_T {\tilde
r}_{\rm out}{\tilde t}\,\rangle$ increase as $k_T$ increases.  At small $k_T$, the values of $\langle 
\beta_T^2{\tilde t}^2 \rangle$ are larger than those of $2\langle \beta_T {\tilde r}_{\rm out} {\tilde 
t}\,\rangle$.  But at large $k_T$ the values of $2\langle \beta_T {\tilde r}_{\rm out} {\tilde t}\,\rangle$ 
are close even larger than the corresponding results of $\langle\beta_T^2{\tilde t}^2\rangle$.  
Both $\langle\beta_T^2{\tilde t}^2
\rangle$ and $2\langle \beta_T {\tilde r}_{\rm out} {\tilde t}\,\rangle$ decrease with collision centrality.
The near values of $\langle\beta_T^2{\tilde t}^2\rangle$ and $2\langle\beta_T{\tilde r}_{\rm out}{\tilde t}\,
\rangle$ at a fixed $k_T$ lead to the HBT results $R_{\rm out}/R_{\rm side}\approx 1$ for the granular sources.
From Fig. \ref{F-rort} (e) and (f) one can see that the average transverse velocities of the pair at the RHIC
energy are higher than those at the LHC energy at large $k_T$.  However, the results of the average longitudinal
velocities are opposite.  For a fixed $k_T$, a larger $\beta_L$ means a larger $k_L$, and therefore a larger
$E_k$ ($E_k^2=k_T^2+k_L^2+m_k^2$) and smaller $\beta_T$.  It will be seen that the reason for the larger
${\overline \beta}_L$ at large $k_T$ at the LHC energy is the larger longitudinal expansion of the granular
source at the LHC energy as compared to that at the RHIC energy (see Fig. \ref{F-tsv} (c) and (d)).

In Fig. \ref{F-tsv}, we show the average emission time ${\overline t}=\langle\,t\,\rangle (k_T)$, the
time variance root $\Delta t=[\langle {\tilde t}^2\rangle]^{1/2}(k_T)$, and the average source transverse
and longitudinal velocities ${\overline v}_{_{ST}}(k_T)$ and ${\overline v}_{_{SL}}(k_T)$ for the granular
sources at the RHIC and LHC energies.  The error bars for the results are statistical errors.  One can see
that the average emission time decreases with increasing collision centrality and increases with increasing collision energy.
It is because that the average droplet velocity increases with decreasing collision centrality and
increasing collision energy.  The larger the droplet velocity, the larger time delay is.  The values
of $\Delta t$, which is also referred as the source lifetime, are almost independent of $k_T$ and smaller
as compared to the values of ${\overline t}$.  For a fixed $k_T$, the average source velocities increase
with decreasing collision centrality.  At large $k_T$, the average longitudinal velocities of the granular
sources at the LHC energy are larger than the corresponding values at the RHIC energy.  It is because that
the pion pairs with larger $k_T$ at the LHC energy correspond to a larger average emission time in the
source center-of-mass frame (see Fig. \ref{F-tsv} (b)), and therefore have larger average longitudinal
source velocities as compared to those at the RHIC energy (see Fig. \ref{F-vt} (b) and (d)).  This result
is consistent with the results of ${\overline \beta}_L$ shown in Fig. \ref{F-rort} (e) and (f).  The large
longitudinal velocity of the granular source at the LHC energy boosts strongly the pair momentum of the
particles with almost the same emission direction, and leads to the large ${\overline \beta}_L$ results.

Finally, it should be noted that the emission time mentioned in the paper is the time recorded from
the initial state of the granular source.  The real emission time from the beginning of collision
should also plus the system pre-equilibrium time $\tau_0$ and the breakup time $t_0$, which should be
different for different collision energies \cite{WNZ11} and centralities.  However, the lifetime of the
granular source is independent of the time original point.  The small source lifetime is a character of
the granular source \cite{WNZ04,WNZ06,WNZ07,WNZ11,WNZ11J}.

\section{Summary and Discussion}

We systemically investigate the pion transverse momentum spectrum, elliptic flow, and HBT interferometry
in the granular source model for the heavy ion collisions at the RHIC highest energy and the LHC energy.
The centrality and energy dependence of the observables at the two energies are examined.  By comparing
the granular source results with the experimental data of the Au-Au collisions at $\sqrt{s_{NN}}=200$
GeV at the RHIC and the Pb-Pb collisions at $\sqrt{s_{NN}}=2.76$ TeV at the LHC with different collision
centralities, we investigate the geometry and velocity parameters in the granular source model as a
function of collision centrality and energy.  The space-time structure and expansion velocities of the
granular sources at the RHIC and LHC energies with different centralities are examined.  Our investigations
indicate that the granular source model can well reproduce the experimental data of pion transverse momentum
spectra, elliptic flow, and HBT radii in the Au-Au collisions at $\sqrt{s_{NN}}=$ 200 GeV with 0--5\%,
10--20\%, and 30--50\% centralities \cite{STA-spe04z,PHE-spe04z,STA-v2-05z,STA-hbt05z}, and in the Pb-Pb
collisions at $\sqrt{s_{NN}}=$ 2.76 TeV with 10--20\% and 40--50\% centralities
\cite{ALI-spe13z,ALI-v2-11z,ALI-hbt11z}.
The experimental data of pion momentum spectra, elliptic flow, and HBT radii impose very strict constraints
on the parameters in the granular source model.  They exhibit certain regularities for collision centrality
and energy.  The space-time structure and expansion velocities of the granular source are consistent with
that reflected by the observables.

In the granular source model we assume that the system created in the ultrarelativistic heavy ion
collisions occurs fragmentation and forms the granular source of QGP droplets due to the dynamical
instability in the fast expansion at the early stage and the surface tension of the strongly coupled
QGP.  We use ideal hydrodynamics to describe the droplet evolution and assume a Gaussian distribution
for the droplet radius for simplicity.  Because many droplets evolve simultaneously, the source
lifetime is smaller as compared to that of a continued big source, which evolves in whole and freeze
out from the source surface.  For the granular source, the distribution of particle-emitting points
of single event presents a clump-structure during the source evolution.  However, the distribution
of source points for many events is continued and presents a volume distribution because of the
contribution from the droplets in the central region of the source.  The short source lifetime,
clump-structure of source points distribution of single event, and volume distribution of source
points for many events are the characters of the granular source.

The investigations for the granular source parameters indicate that the QGP droplets initially
distribute in larger transverse and longitudinal regions for the more central and higher energy
collisions.  So, the distribution width of source points increases with increasing collision 
centrality and energy.  In transverse direction, the droplet velocity of central collision is 
smaller than that of peripheral collision in the center region of the source, although the droplet 
transverse velocity of central collision is larger at the edge of the source.  The average 
longitudinal velocity of droplet is larger than the average transverse velocity of droplet.  The 
droplet longitudinal velocity increases with decreasing collision centrality.  The larger droplet
velocities in the peripheral collisions lead to larger average velocities of source.  Because
the difference of the droplet transverse velocities in and out of reaction plane decreases
with collision centrality, the elliptic flow decreasing with collision centrality.  In HBT
interferometry, the difference between the transverse HBT radii $R_{\rm out}$ and $R_{\rm side}$
is related to the transverse velocity of particle pair, the source lifetime, and the space-time
correlation of source points.  Both the quantities $\langle {\beta_T \,\tilde t}^2 \rangle$ and
$2\langle {\beta_T \, \tilde r}_{\rm out} {\tilde t} \, \rangle$ increase with the decreasing
collision centrality and increasing transverse momentum of the pair, $k_T$.  However, the
difference of the two quantities for a fixed $k_T$ is approximately equal to zero.  This leads
to the results of $R_{\rm out}(k_T)\approx R_{\rm side}(k_T)$.

While one may argue on the details and the number of parameters used in the granular source model,
the consistent explanation of a large number of measured one-particle and two-particle correlation
quantities suggests that this model description captures some fundamental features of the space-time
dynamics.  After all, it is a challenge to a model to explain the experimental data of the momentum
spectra, elliptic flow, and HBT radii simultaneously at the RHIC and LHC energies, and the final
criterion for a model is experiments.  It will be of interest to improve the granular source model
and investigate the effects of QGP viscosity and droplet interaction on the granular source parameters.
On the other hand, the studies of the forming mechanism and signals of the granular source will be of
interest. \\

\begin{acknowledgments}
We thank Dr. L. Cheng, Dr. U. Heinz, Dr. S. Jeon, and Dr. H. C. Song for helpful discussions.  This
research was supported by the National Natural Science Foundation of China under Grant No. 11275037.
\end{acknowledgments}

\end{document}